# Performance Evaluation of on demand and Table driven Protocol for Wireless Ad hoc Network

**PATIL V.P.**

**Department of Electronics and Telecommunication Engineering.**
**Smt. Indira Gandhi college of Engineering, New Mumbai (INDIA).**
**bkvpp@rediffmail.com.**

**Abstract:** Mobile Ad-Hoc Network (MANET) is a wireless network without infrastructure. It is a kind of wireless ad-hoc network, and is a self configuring network of mobile routers connected by wireless links. The routers are free to move randomly and organize themselves arbitrarily, thus the network's wireless topology may change rapidly and unpredictably. Such a network may operate in a standalone fashion, or may be connected to the larger Internet. There are various routing protocols available for MANETs. The most popular ones are DSR, AODV and DSDV. This paper examines two routing protocols for mobile ad hoc networks– the Destination Sequenced Distance Vector (DSDV), the table- driven protocol and the Ad hoc On- Demand Distance Vector routing (AODV), an On –Demand protocol and evaluates both protocols based on packet delivery fraction ,average end to end delay, throughput and routing overhead while varying pause time. The performance evaluation has been done by using simulation tool NS2 which is the main simulator, NAM (Network Animator) and excel graph which is used for preparing the graphs from the trace files. Simulation revealed that although DSDV perfectly scales to small networks with low node speeds, AODV is preferred due to its more efficient use of bandwidth.

## I. Introduction

A Mobile Ad hoc Network (MANET) [1,2] is a kind of wireless ad-hoc network, and is a self-configuring network of mobile routers (and associated hosts) connected by wireless links – the union of which forms an arbitrary topology. The routers are free to move randomly and organize themselves arbitrarily; thus, the network's wireless topology may change rapidly and unpredictably. Such a network may operate in a standalone fashion, or may be connected to the larger Internet.

Routing protocol in MANET can be classified into several ways depending upon their network structure, communication model, routing strategy, and state information and so on but most of these are done depending on routing strategy and network structure. Based on the routing strategy the routing protocols can be classified into two parts: 1.Table driven and 2. Source initiated (on demand) while depending on the network structure these are classified as flat routing, hierarchical routing and geographic position assisted routing. Flat routing covers both routing protocols based on routing strategy [6]. In this paper two ad hoc routing protocols are used, AODV and DSDV. AODV is Reactive (On demand) where as DSDV is Proactive (Table driven) Routing protocol.

The following list of challenges shows the inefficiencies and limitations that have to be overcome in a MANET environment [4]:

a) Routing Overhead: In wireless ad hoc networks, nodes often change their location within network. So, some stale routes are generated in the routing table which leads to unnecessary routing overhead.

b) Limited wireless transmission range: In wireless networks the radio band will be limited and hence data rates it can offer are much lesser than what a wired network can offer. This requires the routing protocols in wireless networks to use the bandwidth always in an optimal manner by keeping the overhead as low as possible [3].

c) Battery constraints: Devices used in these networks have restrictions on the power source in order to maintain portability, size and weight of the device. By increasing the power and processing ability makes the nodes bulky and less portable. So only MANET nodes has to optimally use this resource [5].

d) Time-varying wireless link characteristics: The wireless channel is susceptible to a variety of transmission restrictions duo to path loss, fading, interference and blockage. These factors affect the range, data rate, and the reliability of the wireless transmission [3].

e) Packet losses due to transmission errors: Ad hoc wireless networks experiences a much higher packet loss due to factors such as high bit error rate (BER) in the wireless channel, increased collisions due to the presence of hidden terminals, presence of interference, location dependent contention, frequent path breaks due to mobility of nodes, and the inherent fading properties of the wireless channel [3].

f) Broadcast nature of the wireless medium: The broadcast nature of the radio channel, that is, transmissions made by a node are received by all nodes within its direct transmission range. When a node is receiving data, no other node in its neighborhood, apart from the sender, should transmit. A node should get access to the shared medium only when its transmissions do not affect any ongoing session. Since multiple nodes may contend for the channel simultaneously, the possibility of packet collisions is quite high in wireless networks [3].

g) Mobility-induced route changes: The network topology in an ad hoc wireless network is highly dynamic due to the movement of nodes; hence an on-going session suffers frequent path breaks. This situation often leads to frequent route changes. Therefore mobility management itself is very vast research topic in ad hoc networks [5].

h) Security issues: The radio channel used for ad hoc networks is broadcast in nature and is shared by all the nodes in the network. Data transmitted by a node is received by all the nodes within its direct transmission range. So an attacker can easily snoop and modify the data being transmitted in the network. Here the requirement of confidentiality can be violated if an adversary is also able to interpret the data gathered through snooping [3].

## II. Previous Related Work

A Several authors have done the qualitative and quantitative analysis of Ad Hoc Routing Protocols by means of different performance metrics. They have used different simulators for this purpose. Each one of them has tried to improve some network parameters and have some drawbacks.

1) Md.*Anisur Rahman,Md.Shahidual Islam, Alex Televasky* [15] analyzed that Packet dropping rate for DSR is very less than DSDV and AODV indicating its highest efficiency. Both AODV and DSR perform better under high mobility than DSDV. High mobility occurs due to frequent link failures and the overhead involved in updating all the nodes with the new routing information as in DSDV is much more than that involved in AODV and DSR.

2) *B. Cameron Lesiuk*[16] presented an overview of ad hoc routing principles and thereby demonstrating how these differ from conventional routing. Three proposed ad hoc routing protocols, DSDV, TORA, and DSR were presented and analyzed.

3) *N Vetrivelan & Dr. A V Reddy* [18] studied and analyzed the performance differentials using varying network size and simulation times. They performed two simulation experiments for 10 & 25 nodes for simulation time up to 100 sec.

4) *A.E. Mahmoud, R. Khalaf & A, Kayssi*[17] studied & analyzed three protocols AODV, DSDV and I-DSDV & were simulated using NS-2 package and were compared in terms of packet delivery ratio, end to end delay and routing overhead in different environment; varying number of nodes, speed and pause time. Simulation results show that I-DSDV compared with DSDV, it reduces the number of dropped data packets with little increased overhead at higher rates of node mobility but still can't compete with AODV in higher node speed and number of node.

5) *S. Gowrishanker et al* [19] made the Analysis of AODV and OLSR by using NS-2 simulator, the simulation period for each scenario was 900 seconds and the simulated mobility network area was 800 m x 500 m rectangle. In each simulation scenario, the nodes were initially located at the centre of the simulation region. The nodes start moving after the first 10 seconds of simulated time. The application used to generate is CBR traffic and IP is used as Network layer protocol.

6) *Arunkumar B R et al.* [20] presented their observations regarding the performance comparison of the routing protocols for variable bit rate (VBR) in mobile ad hoc networks (MANETs). They perform extensive simulations, using NS-2 simulator [13]. They have concluded that reactive protocols perform better than proactive protocols.

7) *S. P. Setty et.al.*[21] evaluated the performance of existing wireless routing protocol AODV in various nodes placement models like Grid, Random and Uniform using QualNet 5.0.

8) *Khan et al*. [22] studied and compared the performance of routing protocols by using NCTUns 4.0 network simulator. In this paper, performance of routing protocols was evaluated by varying number of nodes in multiples of 5 in the ad hoc network. The simulations were carried out for 70 seconds of the simulation time. The packet size was fixed to 1400 bytes.

*9) Mr.Rafi U Zamam* [25] studied & compared the performance of DSDV, AODV and DSR routing protocols for ad hoc networks using NS-2 simulations. In this paper, it has been observed that the competitive reactive routing protocols, AODV and DSR, both show better performance than the other in terms of certain metrics. It is still difficult to determine which of them has overall better performance in MANET.

10) *Jorg D.O*. [23] studied the performance comparison of different routing protocols on network topology changes resulting from link breaks, node movement, etc. In his paper performance of routing protocols was evaluated by varying number of nodes etc. But he did not investigate the performance of protocols under heavy loads (high mobility +large number of traffic sources + larger number of nodes in the network), which may lead to congestion situations.

11) *J Broch et al*. [24] performed experiments for performance comparison of both proactive and reactive routing protocols. In their Ns-2 simulation, a network size of 50 nodes with varying pause times and various movement patterns were chosen.

12*) C.E. Perkins & P. Bhagwat [27]* proposed an efficient DSDV (Eff-DSDV) protocol for ad hoc networks. Eff-DSDV overcomes the problem of stale routes, and thereby improves the performance of regular DSDV. The proposed protocol has been implemented in the NCTUns Simulator and performance comparison has been made with regular DSDV and DSR protocols. The performance metrics considered are packet-delivery ratio, end-end delay, dropped packets, routing overhead, route length. It has been found after analysis that the performance of Eff-DSDV is superior to regular DSDV and sometimes better than DSR in certain cases.

*13) Vahid Garousi* [26] performed an analysis of network traffic in ad-hoc networks based on the DSDV protocol with an emphasis on mobility and communication patterns of the nodes. In this paper, he observed that simulations measured the ability of DSDV routing protocol to react to multi-hop ad-hoc network topology changes in terms of scene size, mobile nodes movement, number of connections among nodes, and also the amount of data each mobile node transmits.

### III. Protocol Description

**3.1 Ad Hoc on Demand Distance Vector**

The Ad Hoc on Demand Distance Vector (AODV) [28, 29] routing algorithm is a source initiated, on demand driven, routing protocol. Since the routing is "on demand", a route is only traced when a source node wants to establish communication with a specific destination. The route remains established as long as it is needed for further communication. Furthermore, another feature of AODV is its use of a "destination sequence number" for every route entry. This number is included in the RREQ (Route Request) of any node that desires to send data. These numbers are used to ensure the "freshness" of routing information. For instance, a requesting node always chooses the route with the greatest sequence number to communicate with its destination node. Once a fresh path is found, a RREP (Route Reply) is sent back to the requesting node. AODV also has the necessary mechanism to inform network nodes of any possible link break that might have occurred in the network [7, 8, 30].Thus in short it consist of two phases: Route request and route maintenance.

**3.2 DSDV**

DSDV is one of the most well known table-driven routing algorithms for MANETs. The distance vector algorithm works on a classical Distributed Bellman-Ford (DBF) algorithm [8, 12]. DSDV is a distance vector algorithm which uses sequence numbers originated and updated by the destination, to avoid the looping problem caused by stale routing information. In DSDV, each node maintains a routing table which is constantly and periodically updated (not on-demand) and advertised to each of the node's current neighbors. Each entry in the routing table has the last known destination sequence number. Each node periodically transmits updates, and it does so immediately when significant new information is available. The data broadcasted by each node will contain its new sequence number and the following information for each new route: the destination's address the number of hops to reach the destination and the sequence number of the information received regarding that destination, as originally stamped by the destination. No assumptions about mobile hosts maintaining any sort of time synchronization or about the phase relationship of the update periods between the mobile nodes are made. Following the traditional distance-vector routing algorithms, these update packets contain information about which nodes are accessible from each node and the number of hops necessary to reach them. Routes with more recent sequence numbers are always the preferred basis for forwarding decisions. Of the paths with the same sequence number, those with the smallest metric (number of hops to the destination) will be used. The addresses stored in the route tables will correspond to the layer at which the DSDV protocol is operated. Operation at layer 3 will use network layer addresses for the next hop and destination addresses, and operation at layer 2 will use layer-2 MAC addresses [9, 10, 11, 12].

## IV. Experimental Setup Using Simulation

Both routing techniques were simulated in the same environment using Network Simulator (ns-2) [13]. Both AODV and DSDV were tested for different network performance parameters and by varying the pause time. The algorithms were tested using 50 nodes. For each protocol, we investigated four performance criteria:

i) Throughput, ii) Routing Overhead, iii) Packet loss ratio, iv)Routing overhead.

*Packet delivery ratio:* The packet delivery ratio in this simulation is defined as the ratio between the number of packets sent by constant bit rate sources (CBR,"application layer") and the number of received packets by the CBR sink at destination. It describes percentage of the packets which reach the destination.

*Routing Overhead:* It is the number of packet generated by routing protocol during the simulation. The generation of an important overhead will decrease the protocol performance.

*Average end-to-end delay of data packets:* There are possible delays caused by buffering during route discovery latency, queuing at the interface queue, retransmission delays at the MAC, and propagation and transfer times. Once the time difference between every CBR packet sent and received was recorded, dividing the total time difference over the total number of CBR packets received gave the average end-to-end delay for the received packets. This metric describes the packet delivery time: the lower the end-to-end delay the better the application performance [6].

Table 1: Simulation setup

| Platform | Operating system windows xp sp2(crywin 1.7) |
|---|---|
| Ns version | Ns-allinone-2.29 |
| Pause time | 0,20,40,60,80,100 |
| No. of nodes | 50 wireless nodes |
| Simulation time | 200s |
| Traffic type | UDP |
| Transmission range | 250m |
| CBR packet size | 512 packets |
| Simulation area size | 500*500 m |
| Mobility model | Random Waypoint Model |
| Node speed | Fixed to 25 m/s |
| Bandwidth | 2 Mbps |

## V. Simulation Analysis and Performance Measurement

In this section, the network simulation is implemented using the NS-2 simulation tool [13]. While comparing two protocols, we focused on four performance measurements such as Average Delay, Packet Delivery Fraction, throughput, and routing overhead.

**(i) Packet delivery fraction**: The ratio of the number of data packets successfully delivered to the destinations to those generated by UDP sources. Packet delivery fraction = (Received packets/Sent packets)*100. Fig 1 shows a relation comparison between both the routing protocols on the basis of packet delivery fraction as a function of pause time.

As shown in Fig. 1, DSDV results in lower packet drop than AODV. This is due to the extensive routing information exchanged between the nodes at regular intervals providing a correct, up to date route at all times. Also, no additional packet drop is noticed as speed increases, since the routing updates become more frequent, making the packet drop rates almost unaffected. This feature is not present in AODV. Since the routes are only generated upon request, a route may become outdated by the time the route request is generated and the route reply would arrive. The packets transmitted during this transient period run the risk of being dropped by the network.

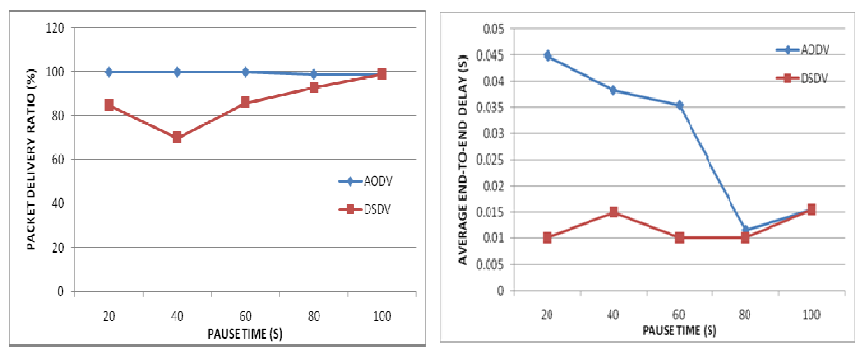

Fig.1: Packet delivery ratio of AODV-DSDV.   Fig 2: Average End-to-End Delay of AODV-DSDV

**(ii) Average End to end delay of data packets**: The average time from the beginning of a packet transmission at a source node until packet delivery to a destination. This includes delays caused by buffering of data packets during route discovery, queuing at the interface queue, retransmission delays at the MAC, and propagation and transfer times. Calculate the send(S) time (t) and receive (R) time (T) and average it. It is seen that average end- to-end delay is less for DSDV protocol than AODV protocol. This is shown in fig. 2.

**iii) Routing Overhead vs. Pause Time**

Since DSDV is less prone to route stability than AODV, we notice from Fig. 3 that routing overhead generated in DSDV is not as affected as that generated by AODV, although that of DSDV is much higher for the reasons explained previously.

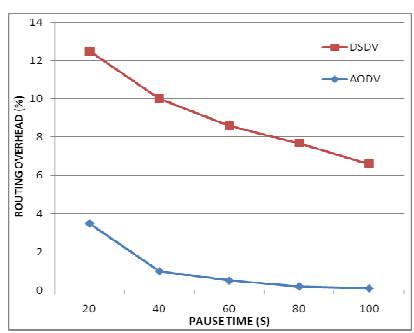 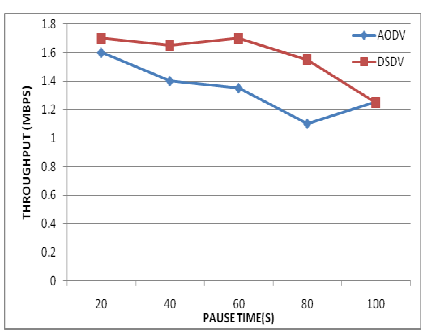

Fig.3: Overhead for AODV and DSDV            Fig.4: Throughput of AODV and DSDV

**iv) Throughput vs. Pause Time**

As shown in Fig 4, AODV and DSDV do not share the same sensitivity to route stability. On one hand, AODV (that operates on a Route Request/Route Reply cycle) shows a significant dependence on route stability, thus its throughput decreases quicker as pause time increases. On the other hand, DSDV (in which the routing tables are periodically updated almost independently of route stability) shows less sensitivity to variations in pause time, which explains the lower decrease rate versus pause time.

## VI. Conclusion

Simulation results show that both of the protocols deliver a greater percentage of the originated data packets when there is little node mobility, converging to 100% delivery ration when there is no node motion. AODV suffers from end to end delays. DSDV packet delivery fraction is very low for high mobility scenarios. We conclude that the AODV protocol is the ideal choice for communication. When the communication has to happen under the UDP protocol as the base, Simulation results reveal that DSDV consumes extensive bandwidth and computation overhead (reaching around 40% in a 30-node network) in the presence of mobility, yielding inferior performance when compared to an on demand algorithm such as AODV (around 12% in a 30-node network). Due to the physical limitations incurred by the medium access control of wireless networks, which physically limits the bandwidth to around 11 Mbps, it is not logical to waste up to 40% of that bandwidth for routing traffic. In terms of throughput, we notice that throughput in AODV is not as affected by speed increase as DSDV. However, DSDV perfectly scales to a small network with low node speeds.